\documentclass[aps,prl,showpacs,preprintnumbers,twocolumn,superscriptaddress]{revtex4-2}
\usepackage{amsmath,amssymb}
\usepackage{bm}
\usepackage{tipa}
\usepackage{upgreek}
\usepackage{comment}
\usepackage{mathrsfs}
\usepackage{graphicx}
\usepackage{braket}
\usepackage{enumitem}
\usepackage{mathbbol}
\usepackage{booktabs}
\usepackage{gensymb}
\usepackage[normalem]{ulem}
\usepackage{color}
\usepackage[colorlinks,bookmarks=true,citecolor=blue,linkcolor=red,urlcolor=blue]{hyperref}
\usepackage{hyperref}

\usepackage{pifont}

\newcommand{\xmark}{\ding{55}}

\begin{document}
	
\title{Electron-phonon  coupling and competing Kekul\'e orders in twisted bilayer graphene}

	\author{Yves H. Kwan}
	\affiliation{Princeton Center for Theoretical Science, Princeton University, Princeton NJ 08544, USA}
 	\author{Glenn Wagner}
	\affiliation{Department of Physics, University of Zurich, Winterthurerstrasse 190, 8057 Zurich, Switzerland}
	\author{Nick Bultinck}
	\affiliation{Rudolf Peierls Centre for Theoretical Physics, Parks Road, Oxford, OX1 3PU, UK}
	\affiliation{Department of Physics, Ghent University, Krijgslaan 281, 9000 Gent, Belgium}
	\author{Steven H. Simon}
	\affiliation{Rudolf Peierls Centre for Theoretical Physics, Parks Road, Oxford, OX1 3PU, UK}
 \author{Erez Berg}
	\affiliation{Department of Condensed Matter Physics, Weizmann Institute of Science, Rehovot, 76100, Israel}
	\author{S.A. Parameswaran}
	\affiliation{Rudolf Peierls Centre for Theoretical Physics, Parks Road, Oxford, OX1 3PU, UK}

\begin{abstract}
Recent scanning tunneling microscopy experiments [K.P. Nuckolls {\it et al.}, arXiv:2303.00024] have revealed the ubiquity of
Kekul\'e charge-density wave order in magic-angle  twisted bilayer graphene.  Most samples are moderately strained and show `incommensurate Kekul\'e spiral' (IKS) order involving a graphene-scale charge density distortion uniaxially modulated on the scale of the moir\'e superlattice, in accord with theoretical predictions. However,  ultra-low strain samples instead show graphene-scale Kekul\'e charge order  that is {\it uniform} on the moir\'e scale. This order, especially prominent near filling factor $\nu=-2$, is unanticipated by theory 
which  predicts a time-reversal {breaking} Kekul\'e \textit{current} order at low strain. We show that including the coupling of moir\'e electrons to
 graphene-scale optical zone-corner (ZC) phonons 
stabilizes a  uniform  Kekul\'e charge ordered state at $|\nu|=2$ with a quantized topological (spin or anomalous Hall) response. Our work clarifies how this phonon-driven selection of electronic order emerges in the strong-coupling regime of moir\'e graphene.

\end{abstract}

\maketitle

\textit{Introduction.---} The interplay of strong electron correlations,  gate-tunable superconductivity, and band topology in `magic-angle' twisted bilayer graphene (MA-TBG) has been the subject of extensive experimental~\cite{Cao2018,Cao2018b,Yankowitz_2019,Lu2019,Park_2021,Zondiner_2020,uri2020mapping,saito2020independent,saito2021isospin,Cao_2021,liu2021tuning,Das_2021,rozen2021entropic,Serlin900,Sharpe_2019,Stepanov_2020,Wu_2021,Saito_2021Hofstadter,nuckolls2023quantum,Grover2022mosaic,Yu2022hofstadter,Yu2022skyrmion,Morisette2022Hunds,Tseng2022nu2QAH,Choi2019,Oh2021unconventional,Nuckolls2020strongly,Xie2021fractional,Diez-Merida2021diode,Jiang2019charge,Arora2020SC,Kerelsky2019,choi2021correlation,Xie2019stm,Wong_2020,pierce2021unconventional} and theoretical investigation~\cite{po2018,xie2020weakfield,XieSub,bultinck_ground_2020,liu2021theories,2020CeaGuinea,Zhang2020HF,ochi2018,KangVafekPRL,Kang2020,vafek2020RG,Liu2021nematic,dodaro2018,TBG4,TBG5,TBG6,SoejimaDMRG,kwan_kekule_2021,PotaszMacDonaldED,zhang2021momentum,klebl2021,shavit2021theory,wu2018phonon,lian2019phonon,wu2019phononlinear,lewandowski2021,Bultinck2019mechanism,hejazi2021,parker2020straininduced,thomson2021,Christos_2020,khalaf2021charged,Chatterjee2020skyrmionic,cea2021electrostatic}.
Although aspects of the phenomenology superficially resemble that of the cuprate high-temperature (high-$T_c$) superconductors,
the nontrivial topology of the eight bands straddling charge neutrality, and the existence of Stoner-like transitions indicative of the formation of flavour-polarized  broken-symmetry states, challenge the  applicability of the Hubbard-type models familiar from high-$T_c$ to the narrow bands in MA-TBG. This  has stimulated a distinct perspective~\cite{bultinck_ground_2020,TBG4} rooted instead in the physics of quantum Hall ferromagnets (QHFM), best  motivated by the approximation of initially ignoring the single-particle dispersion and working in the `chiral limit'~\cite{tarnopolsky_origin_2019} of vanishing interlayer same-sublattice tunneling. In the resulting strong-coupling problem, flavor-polarized insulators  minimize the interaction energy (owing to  Pauli exclusion) at densities of  $|\nu|=0,1,2,3$ electrons per moir\'e unit cell, corresponding to filling $\nu+4$ of the eight central bands. Single-particle terms and realistic interlayer tunneling perturbatively lift the large degeneracy of the  resulting manifold of broken-symmetry states, selecting specific correlated insulators at  integer $\nu$~\cite{bultinck_ground_2020,TBG4,KangVafekPRL}.

\begin{figure*}[t!]
    \centering
    \includegraphics[width=0.85\linewidth]{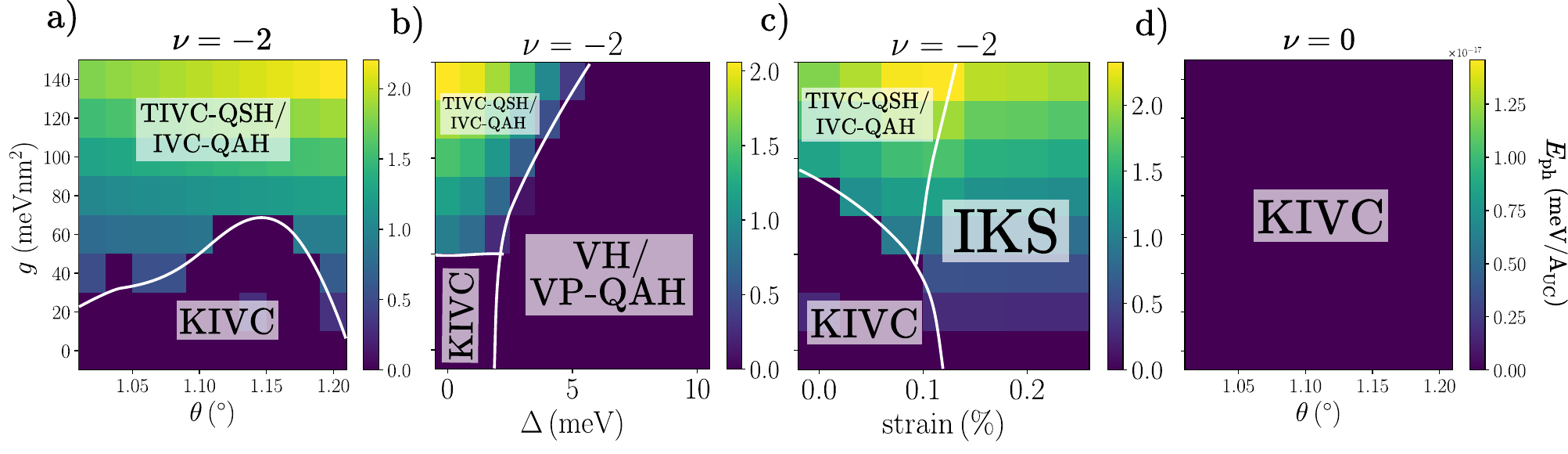}
    \caption{\textbf{Competing orders at even integer filling $\nu$.} Color shows the lattice 
    distortion energy $E_{\text{ph}}$; Kekul\'e charge density order is present in the electronic sector for $E_{\text{ph}}\neq0$. White lines show approximate phase boundaries. All phase diagrams computed in the graphene scheme. a,d) Phase diagram of electron-phonon coupling $g$ vs twist angle $\theta$ for $|\nu|=0,2$ respectively. $w_\text{AA}=60\,\text{meV},w_\text{AB}=110\,\text{meV}$. b) Dependence on sublattice mass $\Delta\sigma_z$ applied to the bottom layer. $\theta=1.10^\circ,w_{\text{AA}}=50\,\text{meV}$. c) Dependence on heterostrain with strength $\epsilon$, with strain axis along $\hat{x}$. $\theta=1.12^\circ,w_\text{AA}=77\,\text{meV}$. [QAH: quantized anomalous Hall, KIVC: Kramers intervalley coherent, TIVC: time-reversal IVC , VP: valley polarized, QSH: quantum spin Hall]}
    \label{fig:all_phase_diagram}
\end{figure*}

Strong-coupling approaches predict
gapped insulators at charge neutrality ($\nu=0$) and time-reversal breaking quantized anomalous Hall (QAH) behaviour or stripe order at odd $\nu$~\cite{Kang2020,Xie2023nu3}. This is in  contradiction to many experiments, that often find semimetallic behaviour $\nu=0$~\cite{Park_2021,Cao2018,Cao2018b,Yankowitz_2019,Cao_2021,liu2021tuning,Zondiner_2020,uri2020mapping,saito2020independent,Das_2021,saito2021isospin,rozen2021entropic}, little transport evidence for gaps at $\nu=\pm 1$, and no QAH response at $\nu=\pm3$ unless aligned with a hexagonal boron nitride substrate~\cite{Serlin900,Sharpe_2019}. Spurred by this mismatch of theory and experiment, Ref.~\cite{kwan_kekule_2021} proposed a new type of broken-symmetry order, dubbed the incommensurate Kekul\'e spiral (IKS), as the ground state for MA-TBG  at intermediate coupling. For modest uniaxial heterostrains~\cite{Kerelsky2019,choi2021correlation,Xie2019stm,Wong_2020} (where layers are strained relative to each other) sufficient to stabilize a semimetal at neutrality~\cite{parker2020straininduced}, 
IKS is the unique Hartree-Fock ground state for all nonzero integer $|\nu|<4$. It exhibits a clear gap and vanishing QAH response for $|\nu|=2, 3$, and is gapless for $|\nu|=1$, consistent with most transport experiments. IKS order also persists to finite doping away from integer $\nu$, seeding Fermi surfaces~\cite{wagner_global_2021}  consistent with Landau fans observed in magnetotransport~\cite{Sharpe_2019,Serlin900,Lu2019,Cao2018,Cao2018b,Yankowitz_2019,Park_2021,Stepanov_2020,Wu_2021,Zondiner_2020,uri2020mapping,saito2020independent,Saito_2021Hofstadter,saito2021isospin}. Most strikingly,  IKS  involves a specific graphene-scale Kekul\'e charge density order that triples the graphene unit cell but is slowly modulated on the moir\'e scale. This   multiscale spatial symmetry-breaking  is a sharp signature of IKS order, recently used to diagnose its presence in MA-TBG via  scanning tunneling microscopy (STM) ~\cite{nuckolls2023quantum}.

These experiments find robust  IKS order in samples with modest strain, with a period 
 of approximately three moir\'e unit cells, in excellent agreement with theoretical predictions. However, at $\nu=-2$ where the data are most extensive, ultra-low-strain samples  show
 Kekul\'e charge order that also triples the graphene unit cell, but is at $q=0$, i.e. unmodulated on the moir\'e scale, in sharp contrast to IKS. This  contradicts  the near-unanimous theoretical prediction in this limit of a $q=0$
 Kekul\'e {\it current} order, dubbed the ``Kramers-Intervalley Coherent" (KIVC) state, whose STM signature vanishes by symmetry~\cite{Princeton_STM_Theory,Berkeley_STM_Theory}. Instead, the observed order  resembles the so-called TIVC state (`T' denotes a spinless implementation of time-reversal), which can be viewed roughly as a charge counterpart of KIVC. However, it is unclear why TIVC becomes a competitive  ground state at low strain.

Here, we show that electron-phonon coupling (EPC) provides a natural explanation for the emergence of low-strain TIVC order. To wit, the zone-corner in-plane optical phonon modes 
--- which link the microscopic valleys --- couple strongly to the Kekul\'e density distortion, lowering the energy of  TIVC relative to   KIVC. The competition is particularly transparent at strong coupling, where  EPC generates a new term in the anisotropic nonlinear sigma model (NLSM) that describes  selection between distinct $q=0$ insulators. This  clarifies that while small relative to the bare Coulomb scale, EPC is comparable in strength to other perturbations that move away from strong coupling. We therefore perform numerical Hartree-Fock  (HF) simulations (Fig.~\ref{fig:all_phase_diagram}) to capture this competition in the regime of intermediate coupling that appears on leaving the chiral-flat limit by tuning interlayer tunneling, strain, and twist angle. Our work shows that the strong-coupling phase structure at low strain is more nuanced than previously thought, and identifies a key role for phonons in selecting between competing interaction-driven ordered states.

\textit{Model.---} We study the Hamiltonian $\hat{H}_\text{tot}=\hat{H}_\text{BM}+\hat{H}_\text{int}+\hat{H}_\text{EPC}+\hat{H}_{A_1}$. Here, $\hat{H}_\text{BM}$ is the standard single-particle Bistritzer-MacDonald (BM) model~\cite{Bistritzer2011} that depends on the twist angle $\theta$ and sublattice-dependent hopping matrix elements $w_\text{AB}=110\,$meV and $w_\text{AA}$, which we will tune starting from the chiral limit $w_{\text{AA}} =0$. $\hat{H}_\text{int}$ describes dual-gate screened Coulomb interactions $V(q)=\frac{e^2}{2\epsilon_0\epsilon_r q}\tanh qd$, with screening length $d=25\,$nm and permittivity $\epsilon_r=10$. To avoid double-counting, we choose the zero of interactions to correspond to the density of two decoupled graphene layers (the so-called `graphene' subtraction scheme), though we investigate alternatives in Ref.~\cite{SupMat}.
$\hat{H}_\text{ph}=\hbar\omega\sum_{l\alpha\bm{q}}\hat{a}^\dagger_{l\alpha}(\bm{q})\hat{a}^{\phantom{\dagger}}_{l\alpha}(\bm{q})
$
describes  graphene zone-corner (ZC) in-plane transverse optical phonons $A_1,B_1$, which couple  to continuum electrons in each layer  via~\cite{Wu2018EPC,Basko2008EPC,Chatterjee2020skyrmionic, Angeli2019JT,Blason2022HFphonon}
\begin{equation}\label{eq:HEPC}
\begin{gathered}
\hat{H}_\text{EPC}=\mathcal{F}\sum_{l\alpha}\int_{\bm{r}}\hat{\psi}_l^\dagger(\bm{r})\left[\hat{u}_{l\alpha}(\bm{r})\Gamma_{\alpha}\right]\hat{\psi}_l^{\phantom{\dagger}}(\bm{r})
\end{gathered}
\end{equation}
with $\hat{u}_{l\alpha}(\bm{r})=\mathcal{D}\sum_{\bm{q}}e^{i\bm{q}\cdot\bm{r}}\left[\hat{a}_{l\alpha}(\bm{q})+\hat{a}^\dagger_{l\alpha}(-\bm{q})\right]$. Here,  $\hat{\psi}_l(\bm{r})$ is a spinor in spin ($s$), valley ($\tau$) and sublattice ($\sigma$) space, $\mathcal{F}, \mathcal{D}$ absorb various phonon parameters, and we approximate the phonon dispersion $\hbar\omega\simeq 160\,$meV as constant (since the optical mode is roughly flat within the BM model cutoff). Each layer $l$ has two degenerate ZC modes $\alpha=a,b$ with intervalley coupling matrices $\Gamma_a=\tau_x\sigma_x,\Gamma_b=\tau_y\sigma_x$. We define a characteristic (dimensionful)  coupling 
\begin{equation}
    g=A\frac{\mathcal{F}^2\mathcal{D}^2}{\hbar\omega},
\end{equation}
where $A$ is the system area. Typical estimates put $g\simeq 70\,$meVnm$^2$~\cite{Wu2018EPC}, but as these can vary widely~\cite{Basko2008EPC}, we will view it
as a tuning parameter. 
$\hat{H}_\text{tot}$ is invariant under spinless time-reversal $\hat{\mathcal{T}}=\tau_x\mathcal{K}$, $U(1)_V$ valley rotations, $SU(2)_s$ spin rotations, and $D_6$ point-group symmetry.

Strong electron-electron interactions lead to closely competing candidate ground states. Treating $\hat{H}_\text{EPC}$ at mean-field level, the phonons will experience a linear bias term $\sim \text{tr}\,\Gamma_\alpha P$, where $P$ is the electron density matrix, and lower their energy by shifting their vacuum.
The resulting 
energy gain from lattice distortion $E_{\text{ph}}\sim g\left|\text{tr}\,\Gamma_\alpha P\right|^2\geq 0$ is quadratic in $P$, and, crucially, depends on the pattern of flavor symmetry-breaking. For the 
ZC phonons of interest here, this effect is only operative for certain forms of $U(1)_V$-breaking intervalley coherence (IVC). Due to large gaps to the dispersive remote bands, the relevant electronic ordering is concentrated in the central bands. Hence in the following, we consider $\hat{H}_\text{tot}$ projected to the flat bands. Expressions for $E_{\text{ph}}$ in the projected theory are given in Ref.~\cite{SupMat}.

\begin{figure}[t]
    \centering
    \includegraphics[width=1\linewidth]{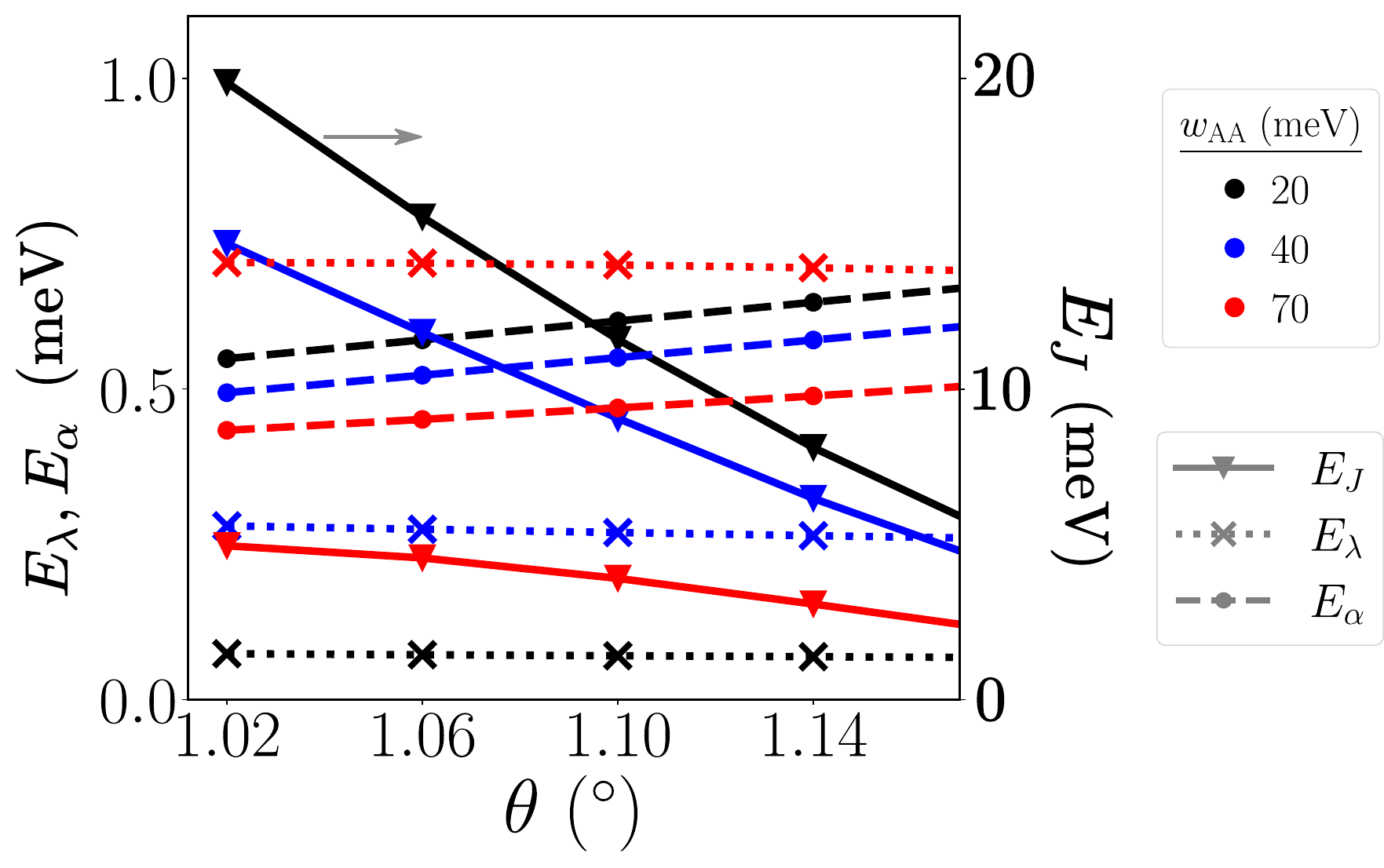}
    \caption{\textbf{Sigma model energy scales.} Parameters of the strong-coupling sigma model (energies are per moir\'e unit cell) for $g=70\,$meVnm$^{2}$ in the graphene scheme. $E_\alpha=\alpha A_{\text{UC}}$ (dashed circles) measures twice the lattice distortion energy per unit cell for a single intervalley coherent Chern band. Note the different scale for $E_J$ (solid triangles).}
    \label{fig:Jlamb}
\end{figure}

\textit{Strong-coupling limit and NLSM.---} To understand the qualitative impact of EPC on ground state selection, we first consider a non-linear sigma model (NLSM) description~\cite{khalaf2021charged,khalaf2020soft,Kwan2022skyrmions}. In the chiral-flat limit with $\kappa=\frac{w_\text{AA}}{w_\text{AB}}=0$ and vanishing bandwidth, we can rotate to the Chern basis which is sublattice polarized and has Chern numbers $C=\tau_z\sigma_z$ (the polarization is imperfect for $\kappa\neq0$). At integer $\nu \equiv  \nu_+ +\nu_- - 4$, the exact ground states are Slater determinants constructed by filling $\nu_+$ bands with $C=1$  and $\nu_-$ bands with $C=-1$, allowing arbitrary rotations within each Chern sector. These `generalized ferromagnets' spontaneously break the $U(4)\times U(4)$ symmetry of the chiral-flat limit to $U(\nu_+)\times U(4-\nu_+)\times U(\nu_-)\times U(4-\nu_-)$, which sets the NLSM target space. Deviations from the chiral-flat limit {\it explicitly} break  $U(4)\times U(4)$, leading to
anisotropies in the NLSM energy density
\begin{equation}\label{eq:NLSM}
\begin{aligned}
    \mathcal{E}[Q]&=\frac{J}{4}\text{tr}\,(Q\gamma_x)^2-\frac{\lambda}{4}\text{tr}\,(Q\gamma_x\eta_z)^2\\
    &\quad-\frac{\alpha}{8}\left[\left(\text{tr}\,Q\eta_x\right)^2+\left(\text{tr}\,Q\eta_y\right)^2\right],
\end{aligned}
\end{equation}
where we have defined the Pauli triplets $\gamma_{x,y,z}=(\sigma_x,\sigma_y\tau_z,\sigma_z\tau_z),\quad \eta_{x,y,z}=(\sigma_x\tau_x,\sigma_x\tau_y,\tau_z)$. $Q=\text{diag}(Q^+,Q^-)$ is the $8\times8$ single-particle density matrix. $Q$ is block-diagonal in Chern sectors $C=\pm$, which satisfies $Q^2=1$ and $\text{tr}\,Q=2\nu$. $J$ and $\lambda$ are previously-computed~\cite{bultinck_ground_2020,Kwan2022skyrmions} terms arising from inter-Chern tunneling (superexchange) and finite $\kappa$ respectively. The $\alpha$ term is new, and represents the phonon energy from coupling to the Kekul\'e charge density. We argue that its inclusion is necessary since its magnitude is comparable to the other anisotropies (Fig.~\ref{fig:Jlamb}).

\begin{table*}
\centering
\newcommand{\colskip}{\hskip 0.15in}
\renewcommand{\arraystretch}{1.15}
\begin{tabular}{
l @{\hskip 0.3in} 
c @{\colskip} 
c @{\colskip} 
c @{\colskip} 
c @{\colskip} 
c @{\colskip} 
c  }\toprule[1.3pt]\addlinespace[0.3em]
Phase & 
$|\nu|$ & 
$\ket{\psi}$ & 
$E$ & 
Kekul\'e & C &
spin pol. 
\\ \midrule
KIVC & 0 & $\left(\ket{KA}+\ket{\bar{K}B})(\ket{KB}-\ket{\bar{K}A}\right)$ & $-2J-2\lambda$ & current & 0 & 0\\
TIVC & 0 & $\left(\ket{KA}+\ket{\bar{K}B})(\ket{KB}+\ket{\bar{K}A}\right)$ & $2J+2\lambda-8\alpha$ & charge &0 & 0\\
VH & 0 & $\ket{KA}\ket{\bar{K}A}$ & $-2J+2\lambda$ & \xmark &0 & 0\\
KIVC & 2 & $\left(\ket{KA\uparrow}+\ket{\bar{K}B\uparrow})(\ket{KB\uparrow}-\ket{\bar{K}A\uparrow}\right)$ & $-2\lambda$ & current& 0 & *\\
TIVC-SP & 2 & $\left(\ket{KA\uparrow}+\ket{\bar{K}B\uparrow})(\ket{KB\uparrow}+\ket{\bar{K}A\uparrow}\right)$ & $2J-2\alpha$ & charge& 0 & 2\\
TIVC-QSH & 2 & $\left(\ket{KA\uparrow}+\ket{\bar{K}B\uparrow})(\ket{KB\downarrow}+\ket{\bar{K}A\downarrow}\right)$ & $-2\alpha$ & charge& 0 & 0\\
IVC-QAH & 2 & $\left(\ket{KA\uparrow}+\ket{\bar{K}B\uparrow})(\ket{KA\downarrow}+\ket{\bar{K}B\downarrow}\right)$ & $-2\alpha$ & charge& 2 & 0\\
VH & 2 & $\ket{KA\uparrow}\ket{\bar{K}A\uparrow}$ & $0$ & \xmark& 0 & *
\\\bottomrule[1.3pt]
\end{tabular}
\caption{\textbf{Valley-unpolarized strong-coupling states at even integer filling.} All states are moir\'e translation symmetric. For neutrality, the state in the $\ket{\psi}$ column is repeated for both spin projections. $E$ denotes energy in the non-linear sigma model. `Current' in Kekul\'e column can refer to both charge and spin currents. $*$ in spin polarization column indicates a degenerate manifold of states obtained by performing $SU_S(2)$-rotation on a subset of the Hilbert space. The corresponding density matrices $Q$ are shown in Ref.~\cite{SupMat}. VH: valley Hall.\label{tab:even}}
\end{table*}

In Tab.~\ref{tab:even}, we list the candidate strong-coupling states at even integers
focusing on uniform valley-unpolarized orders. Prior theory has consistently favored the KIVC at even $\nu$ on the grounds that it maximally satisfies both $J$ and $\lambda$ terms~\cite{bultinck_ground_2020,TBG4}. However, despite its IVC, it does not benefit from EPC. This is due to its $\hat{\mathcal{T}}'=\tau_y\mathcal{K}$ symmetry, which forces the Kekul\'e charge density to vanish~\cite{Princeton_STM_Theory,Berkeley_STM_Theory}. The TIVC is usually ignored due to its energy penalty under $J$ and $\lambda$, but it can gain from the $\alpha$-term since the Kekul\'e densities of its bands interfere constructively. This could be anticipated from the phonon coupling matrices $\eta_x,\eta_y$ in Eq.~\ref{eq:NLSM} which are precisely the order parameters of the TIVC. For finite chiral ratio, this effect decreases as the sublattice polarization of the Chern bands is reduced, but $\alpha$ remains appreciable (Fig.~\ref{fig:Jlamb}a).

For $|\nu|=2$ which is of most relevance to Ref.~\cite{nuckolls2023quantum}, the case for TIVC is strongest, due to nontrivial interplay with spin physics. The prevailing theoretical expectation for the ground state is the spin-polarized KIVC (upto $SU_K(2)\times SU_{\bar{K}}(2)$ spin rotations). The $J$ and $\lambda$ terms are antagonistic towards the spin-polarized TIVC. However by flipping the spin in one Chern sector, we can construct instead a new state, the TIVC-QSH, which does not incur the energy cost $J$ (Tab.~\ref{tab:even}). This is because inter-Chern tunneling is no longer Pauli-blocked, allowing superexchange.  Therefore, it suffices only that $\alpha>\lambda$
for this phase with Kekul\'e charge order to emerge; from Fig.~\ref{fig:Jlamb} we see that these are indeed comparable. Note that the TIVC-QSH satisfies spinful time-reversal and is a quantum spin Hall insulator (hence the name).  Applying spinless $\hat{\mathcal{T}}$ on one spin species produces a degenerate IVC order with $|C|=2$ quantized anomalous Hall (QAH) response,  that we dub the IVC-QAH~\cite{Blason2022HFphonon}. Both this and the TIVC-QSH exhibit quantized topological responses.

At neutrality, the superexchange mechanism that stabilizes TIVC-QSH is Pauli-blocked and KIVC dominates TIVC orders due to the large  $J$, inevitably present in the graphene subtraction scheme. However, the choice of scheme  influences the interaction-renormalized bandwidth, with a particularly strong effect on $J$.
In the `average' scheme~\cite{SupMat}, $J=\lambda=0$ at the magic angle in the chiral-flat limit, ensuring  Kekul\'e charge order for any $\alpha>0$~\cite{SupMat}; away from this limit, $J$ remains small, suggesting a qualitatively different $\nu=0$ phase diagram.

\textit{Hartree-Fock Results.---} To study the phase competition beyond the NLSM, we perform HF calculations for a realistic chiral ratio.  Phonons are included self-consistently by optimizing over products of electronic Slater determinants and phonon coherent states. We assume moir\'e translation symmetry and diagonal spin structure.
Since our model has approximate particle-hole symmetry~\cite{Zou2018emergent,Song2019PHS,Hejazi2019multiple}, we only show data for $\nu\leq0$.

At $\nu=-2$, the ground state is the KIVC for small values of $g$ (Fig.~\ref{fig:all_phase_diagram}). As expected from its $\hat{\mathcal{T}}'$ symmetry, it has vanishing Kekul\'e charge density and does not couple linearly to the ZC phonons. For larger EPC, we find a first-order transition to a gapped phase with finite Kekul\'e charge order, which either satisfies spinful TRS (TIVC-QSH) or is a $|C|=2$ Chern insulator (IVC-QAH). These are degenerate at HF level, exhibit identical Kekul\'e patterns, and possess nearly perfect IVC across the moir\'e Brillouin zone (mBZ). The fact that  $E_{\text{ph}}\approx 2E_\alpha$ (Fig.~\ref{fig:Jlamb}a) strongly suggests that these states are \emph{quantitatively} similar to those in the NLSM limit.

At $\nu=0$, we find only KIVC order for the same parameter window, with no competing Kekul\'e charge orders.

We also study the phase diagrams for odd $\nu$~\cite{SupMat}. For a single Chern band, the anisotropy that selects valley-polarization over IVC is much smaller than the terms in Eq.~\ref{eq:NLSM}~\cite{TBG4}. Therefore, the transition to states with Kekul\'e charge density occurs for weaker EPC. 

Finally, we comment that in the average scheme~\cite{SupMat} at $\nu=0$, a first-order  KIVC-TIVC transition  with increasing $g$  reappears, whereas the phase boundaries at $|\nu|\geq 2$ are largely unchanged. These observations are consistent with the NLSM discussion above. 

Alignment of MA-TBG to the hBN substrate breaks $\hat{C}_{2z}$ symmetry, and can be modeled via a sublattice mass $\Delta\sigma_z$~\cite{Jung2015hBN,Bultinck2019mechanism,Zhang2019anomalous} (though there are likely additional complicated effects~\cite{Long2022hBN,Cea2020hBN,Mao2021hBN,Shin2021hBNasym,Shi2021hBN,Lin2021hBNmis,Lin2020hBNsym,Kwan2021domain,Grover2022mosaic,Wong2023fractional}). As shown in Fig.~\ref{fig:all_phase_diagram}b for $\nu=-2$, the sublattice bias competes with intervalley coherence, and both IVC orders give way to the valley Hall (VH) phase for modest values of $\Delta$.
This is a smooth crossover between states on the pseudospin Bloch sphere, driven by the sublattice potential.

In the EPC-heterostrain phase diagram at $\nu=-2$ (Fig.~\ref{fig:all_phase_diagram}c), all three types of IVC ordering are present. The band gaps of the moir\'e translation invariant ($q=0$) TIVC and KIVC are rapidly suppressed by strain~\cite{parker2020straininduced,kwan_kekule_2021}, yielding to IKS for small strains typical of most MA-TBG devices. Since the IKS possesses Kekul\'e bond order, it can directly couple to the ZC 
phonons, thereby explaining its relative stability against the TIVC for finite $g$. The IKS can be sharply distinguished from the TIVC by its non-zero $q$, trivial spin Chern number, and strongly inhomogeneous IVC in momentum space~\cite{kwan_kekule_2021,wang2022kekule}.

\textit{Discussion.---} While virtual phonons in MA-TBG have previously been invoked to explain superconductivity~\cite{Wu2018EPC,lian2019phonon,Cea2021SC,wu2019phononlinear,Choi2018phonon,Peltonen2018SC,Lewandowski2021pairing} and resolve spin degeneracies via Hund's coupling~\cite{Chatterjee2020skyrmionic,Lake2022synthesis,Morisette2022Hunds,Lake2021reentrant,Khalaf2022nonunitary}, the role played by zone-corner optical phonons here is special: 
by triggering a physical lattice distortion in response to electronic Kekul\'e charge order, 
 phonons {\it directly} participate in ground state selection.  This modifies the physics to the extent that the TIVC, usually considered the least likely strong-coupling order, can emerge as the ground state at even integer $\nu$. This ``valley Jahn-Teller effect''~\cite{Angeli2019JT}  has been previously studied using HF and projected resonating-valence bond (RVB) wavefunctions~\cite{Blason2022HFphonon}. However, while Ref.~\onlinecite{Blason2022HFphonon} did find  that KIVC becomes unstable to Kekul\'e charge order, it did not consider competition with translational-breaking orders like IKS in the presence of strain or substrate alignment, nor, crucially, did it apply the lens of the strong coupling NLSM as we do here. Hence Ref.~\onlinecite{Blason2022HFphonon}  identified the IVC-QAH state as the only possible alternative to KIVC at $\nu=-2$. In contrast, our  NLSM analysis shows that TIVC-QSH and  IVC-QAH are degenerate (certainly at HF level but possibly beyond). TIVC-QSH  is more consistent with the bulk of experiments, that do not see QAH at $|\nu|=2$, and we find that it is suppressed at moderate strain in favor of IKS, consistent with experiments~\cite{nuckolls2023quantum}. On a more technical level, we note that Ref.~\onlinecite{Blason2022HFphonon} used an unusual subtraction scheme, wherein Kekul\'e charge order appears roughly equally stable at $|\nu|=0,2$, in contrast to our graphene-scheme results where it appears to be weaker at neutrality --- again, in potential agreement with experiments~\cite{nuckolls2023quantum}.

  Beyond offering a resolution to an immediate experimental puzzle~\cite{nuckolls2023quantum}, the emergence of TIVC has ramifications for other aspects of correlation physics in moir\'e graphene. It has been argued that experiments in MA-TBG and twisted trilayer graphene (TTG) indicate pairing between opposite spins and valleys in the superconducting dome commonly observed upon hole doping $\nu=-2$~\cite{Lake2022synthesis}. Both TIVC-QSH and IKS preserve spinful TRS and accommodate such pairing (unlike KIVC or IVC-QAH); it would be interesting to explore this further, perhaps using similar techniques to Ref.~\onlinecite{Blason2022HFphonon}. We note that the spin structure is already established at an energy above the weak Hund's coupling~\cite{Morisette2022Hunds} whose sign, determined by a delicate balance between virtual phonons and intervalley Coulomb scattering, is theoretically difficult to compute~\cite{Chatterjee2020skyrmionic}. 

The non-trivial topology of the TIVC also leads to phenomena distinct from the IKS. Topological spin/pseudospin textures carry electrical charge, and may pair if the energetics are favorable~\cite{khalaf2021charged,Chatterjee2022dmrg,Kwan2022skyrmions}. At $|\nu|=2$, the TIVC-QSH exhibits a quantum spin Hall effect protected by $S_z$ conservation. Crucially, the presence of rough edges does not gap the edge modes, unlike the KIVC where the protecting $\hat{\mathcal{T}}'$-symmetry is broken by intervalley scattering at boundaries~\cite{bultinck_ground_2020}. In addition, the TRS-violating IVC-QAH could explain~\cite{Blason2022HFphonon} experiments which see time-reversal symmetry-breaking at $|\nu|=2$~\cite{Tseng2022nu2QAH,Diez-Merida2021diode}, without an extrinsic substrate coupling.

The optical phonon distortion mechanism proposed here is likely also relevant to mirror-symmetric TTG, whose bands resemble MA-TBG except for extra dispersive Dirac cones. TTG  is phenomenologically similar to MA-TBG, e.g. it also hosts superconducting domes proximate to correlated insulators~\cite{Cao2021TTGPauli,Park2021TTGSC,Hao2021TTGelectric,Shen2022TTGDirac,Liu2022TTGisospin,Kim2021TTGunconventional}. TTG has been observed to form solitons and `twistons'~\cite{Twistons}, suggesting the emergence of locally low-strain regions that are ideal for realizing this phonon-induced selection of electronic order. We speculate that intervalley phonons may also influence the phase structure of correlated moir\'e-less graphene multilayers, that exhibit multiple Fermi surface reconstructions and flavor symmetry-breaking transitions~\cite{Zhou2022BBGSC,Seiler2022BBGcascade,delaBarrera2022BBGcascade,Zhang2023BBGSC,Holleis2023BBGSC,Zhou2021RTGhalf,Zhou2021RTGSC}, though the intralayer intersublattice EPC is suppressed in Bernal-stacked structures~\cite{Choi2021dichotomy}.  

We have focused on optical zone-corner phonons since they directly couple to IVC order. There is a plethora of other phonons not considered here, from optical graphene $\Gamma$ modes to low-energy moir\'e acoustic phonons and phasons~\cite{Koshino2019moirephonon,Ochoa2019phason,Koshino2020phononeffective,Angeli2020phonon,Ochoa2022phonondegradation,Gao2022phononsymmetry,Liu2022moirephonon,Lu2022phononmoire,Miao2023phonontruncated,Cappeulluti2023phononflat,Choi2018phonon,Choi2021dichotomy,Angeli2019JT}. Incorporation of additional terms~\cite{Vafek2023continuum,Kang2023pseudo} in the BM model would be useful to recover the particle-hole symmetry breaking seen in experiments. It may also be interesting to examine the role of ZC EPC in heavy fermion formulations of MA-TBG and TTG~\cite{Song2022heavy,Calugaru2023heavy2,Yu2023heavyTTG}.

\begin{acknowledgements}
\textit{Acknowledgements}.--- We thank Steve Kivelson for useful discussions and A. Yazdani and the authors of Ref.~\cite{nuckolls2023quantum}   for sharing their results with us shortly before their publication. We acknowledge support from  the European Research Council (ERC) under the European Union Horizon 2020 Research and Innovation Programme (Grant Agreement Nos.~804213-TMCS, 757867-PARATOP, and 817799-HQMAT) and from EPSRC grant EP/S020527/1.

\end{acknowledgements}

%

\newpage
\clearpage

\begin{appendix}
\onecolumngrid
	\begin{center}
		\textbf{\large --- Supplementary Material ---\\ Electron-phonon coupling and competing Kekul\'e orders in twisted bilayer graphene}\\
		\medskip
		\text{Yves H. Kwan, Glenn Wagner, Nick Bultinck, Steven H. Simon, Erez Berg, S.A. Parameswaran}
	\end{center}

\section{Electron-phonon coupling in the projected theory}

Following Ref.~\cite{Wu2018EPC}, the electron-phonon coupling (EPC) to optical in-plane graphene zone-corner modes can be written
\begin{gather}\label{eq:EPC}
    \hat{H}_{\text{EPC}}=\mathcal{F}\int_{\bm{r}}\sum_{\alpha}\sum_{sl\sigma\sigma'}\hat{u}_{l\alpha}(\bm{r})\hat{\psi}^\dagger_{K,s, l\sigma}(\bm{r})\Gamma^\alpha_{\sigma,\sigma'}\hat{\psi}_{\bar{K},s,l\sigma'}(\bm{r})+\text{h.c.}\\
    \hat{u}_{l\alpha}(\bm{r})=\mathcal{D}\sum_{\bm{q}}e^{i\bm{q}\bm{r}}\left[\hat{a}_{l\alpha}(\bm{q})+\hat{a}^\dagger_{l\alpha}(-\bm{q})\right]\\
    \mathcal{D}=\sqrt{\frac{\hbar}{2N_gM_C\omega_\alpha}}
\end{gather}
where $N_g$ is the number of graphene unit cells in one layer, $M_C$ is the carbon mass, $\omega$ is the phonon frequency, and we have two types of phonon modes $\alpha=a,b$ per layer, which are degenerate. The integral is over the entire system. The coupling matrices are $\Gamma^a=\sigma^x,\Gamma^b=-i\sigma^x$ (note that we have explicitly written the valley indices in $\hat{H}_{\text{EPC}}$). The h.c.~captures processes where $K$ electrons are scattered to $\bar{K}$.

We now project the electron operators in the EPC to the central bands. Recall the central-band projected position operators
\begin{gather}
    \hat{\psi}^\dagger_{\tau,s,I}(\bm{r})=\sum_{\bm{k},n}\phi^*_{\bm{k}\tau nI}(\bm{r})\hat{d}^\dagger_{\tau,s,n}(\bm{k})\\
    \phi_{\bm{k}\tau n I}(\bm{r})=
    \frac{1}
    {\sqrt{A}}e^{i\bm{k}\bm{r}}\sum_{\bm{G}}e^{i\bm{G}\bm{r}}u_{\tau n I}(\bm{k},\bm{G}),
\end{gather}
where $A$ is the total system area, $\bm{k}$ runs over the mBZ in each valley, and $I=(1\text{A},1\text{B},2\text{A},2\text{B})$. Note that we have not included the fast vector which connects from graphene $\Gamma$ to one of the moir\'e $\Gamma_{\text{M}}$ points near valley $K$---this is because the phonon wavevector $\bm{q}$ is slow on the graphene scale and the phonon operators $\hat{a}$ have already captured the fast Kekul\'e mode. The Bloch functions $\phi$ are normalized to the entire system. The Bloch coefficients hence satisfy $\sum_{\bm{G}I}u_{\tau n I}(\bm{k},\bm{G})u^*_{\tau n' I}(\bm{k},\bm{G})=\delta_{nn'}$.

We now massage the $\bar{K}\rightarrow K$ part of the EPC for one in-plane mode (so we temporarily drop the in-plane mode index $\alpha$) 
\begin{align}
    \hat{h}_l&=\frac{\mathcal{F}_{A_1}\mathcal{D}_{A_1}}{A}\int_{\bm{r}}\sum_{\bm{q}}e^{i\bm{q}\bm{r}}\left[\hat{a}_l(\bm{q})+\hat{a}_l^\dagger(-\bm{q})\right]\\
    &\quad \times\sum_{\bm{k}\bm{k}'nn's}e^{i(\bm{k}'-\bm{k})\bm{r}}d^\dagger_{K s n}(\bm{k})d_{\bar{K} s n'}(\bm{k'})\sum_{\bm{G}\bm{G}'\sigma\sigma'}e^{i(\bm{G}'-\bm{G})\bm{r}}u^*_{Knl\sigma}(\bm{k},\bm{G})\Gamma_{\sigma,\sigma'}u_{\bar{K}n'l\sigma'}(\bm{k}',\bm{G}').
\end{align}
We now take the expectation value $\langle d^\dagger_{Ks n}(\bm{k})d_{\bar{K}s n'}(\bm{k'})\rangle=P_{Kn;\bar{K}n'}(\bm{k},s)\delta_{\bm{k},\bm{k}'}$. The integral over $\bm{r}$ enforces $\bm{q}=\bm{G}_p$, where $\bm{G}_p$ is a moir\'e RLV, and fixes $\bm{G}'=\bm{G}-\bm{G}_p$
\begin{gather}
    \hat{h}_l=\mathcal{F}_{A_1}\mathcal{D}_{A_1}\sum_{\bm{G}_p}\left[\hat{a}_l(\bm{G}_p)+\hat{a}_l^\dagger(-\bm{G}_p)\right]\sum_{\bm{k}nn's}P_{Kn;\bar{K}n'}(\bm{k},s)\Lambda_{\Gamma;n,n',l}(\bm{k},-\bm{G}_p)\\
    \Lambda_{\Gamma;n,n',l}(\bm{k},\bm{G}_p)=\sum_{\bm{G},\sigma\sigma'}u^*_{Knl\sigma}(\bm{k},\bm{G})\Gamma_{\sigma,\sigma'}u_{\bar{K}n'l\sigma'}(\bm{k},\bm{G}+\bm{G}_p)=\Lambda^*_{\Gamma^\dagger;n',n,l}(\bm{k},-\bm{G}_p).
\end{gather}
So $\Lambda$ above is a layer-resolved intervalley version of the usual form factor with sublattice contraction.

Consider the Hamiltonian for a single harmonic mode with linear terms
\begin{equation}
    \hat{H}=\epsilon a^\dagger a+\gamma a+\gamma^* a^\dagger.
\end{equation}
One can work with shifted canonical operators $b^\dagger=a^\dagger+\gamma/\epsilon$, leading to 
\begin{equation}
    \hat{H}=\epsilon b^\dagger b-\frac{|\gamma|^2}{\epsilon}
\end{equation}
so that the shifted vacuum has energy $-|\gamma|^2/\epsilon$.

For the case of the ZC EPC in MA-TBG, we will have a $\gamma_{\alpha}(\bm{G}_p)$ and $\epsilon_{\alpha}(\bm{G}_p)=\hbar\omega$ for every layer-mode $l\alpha$ and RLV (we neglect the dispersion of the phonon on the scale of the BM model cutoff). The linear coefficients are
\begin{gather}
    \gamma_{la}(\bm{G}_p)=\mathcal{F}_{A_1}\mathcal{D}_{A_1}\sum_{\bm{k}nn's}\left[P_{Kn;\bar{K}n'}(\bm{k},s)\Lambda_{\sigma^x;n,n',l}(\bm{k},-\bm{G}_p)+P^*_{Kn;\bar{K}n'}(\bm{k},s)\Lambda^*_{\sigma^x;n,n',l}(\bm{k},\bm{G}_p)\right]\\
    \gamma_{lb}(\bm{G}_p)=\mathcal{F}_{A_1}\mathcal{D}_{A_1}\sum_{\bm{k}nn's}\left[-iP_{Kn;\bar{K}n'}(\bm{k},s)\Lambda_{\sigma^x;n,n',l}(\bm{k},-\bm{G}_p)+iP^*_{Kn;\bar{K}n'}(\bm{k},s)\Lambda^*_{\sigma^x;n,n',l}(\bm{k},\bm{G}_p)\right].
\end{gather}
 The total lattice distortion energy is the sum of contributions over all layer-modes and RLVs.

Define the interaction strength parameter
\begin{equation}
    g_\alpha=\frac{A}{N_g}\left(\frac{\mathcal{F}_\alpha}{\hbar\omega_\alpha}\right)^2\frac{\hbar^2}{2M_C}=A\frac{\mathcal{F}_\alpha^2\mathcal{D}_\alpha^2}{\hbar\omega_\alpha}.
\end{equation}

Then the total  negative semi-definite lattice distortion energy $E_{\text{ph}}$ is (in the main text, we consider the negative of this)
\begin{gather}
    \delta E_a=-\frac{g_{A_1}}{A}\sum_{\bm{G}_pl}\left|\sum_{\bm{k}nn's}P_{Kn;\bar{K}n'}(\bm{k},s)\Lambda_{\sigma^x;n,n',l}(\bm{k},-\bm{G}_p)+P^*_{Kn;\bar{K}n'}(\bm{k},s)\Lambda^*_{\sigma^x;n,n',l}(\bm{k},\bm{G}_p)\right|^2\\
    \delta E_b=-\frac{g_{A_1}}{A}\sum_{\bm{G}_pl}\left|\sum_{\bm{k}nn's}-iP_{Kn;\bar{K}n'}(\bm{k},s)\Lambda_{\sigma^x;n,n',l}(\bm{k},-\bm{G}_p)+iP^*_{Kn;\bar{K}n'}(\bm{k},s)\Lambda^*_{\sigma^x;n,n',l}(\bm{k},\bm{G}_p)\right|^2\\
    E_{\text{ph}}=-2\frac{g_{A_1}}{A}\sum_{\bm{G}_pl}\left(\left|\sum_{\bm{k}nn's}P_{Kn;\bar{K}n'}(\bm{k},s)\Lambda_{\sigma^x;n,n',l}(\bm{k},-\bm{G}_p)\right|^2
    +\left|\sum_{\bm{k}nn's}P^*_{Kn;\bar{K}n'}(\bm{k},s)\Lambda^*_{\sigma^x;n,n',l}(\bm{k},\bm{G}_p)\right|^2\right).
\end{gather}

We now discuss how to incorporate the phonons self-consistently in mean-field theory, by augmenting the standard electronic Hartree-Fock procedure with an additional term. In effect, the variational manifold consists of products of an electronic Slater determinant and phonon coherent states. Consider a general EPC Hamiltonian
\begin{equation}
    \hat{H}=\sum_{\alpha}\omega_\alpha \hat{b}^\dagger_\alpha \hat{b}_\alpha + \sum_{\alpha}\sum_{ij}\hat{b}_\alpha\Gamma_{\alpha;ij}\hat{d}^\dagger_i\hat{d}_j+\text{h.c.},
\end{equation}
where $\hat{d}^\dagger$ are the electronic degrees of freedom. For a given electron projector $P_{ij}=\langle\hat{d}^\dagger_i\hat{d}_j\rangle$, the linear coupling experienced by the phonons is $\gamma_\alpha=\sum_{ij}\Gamma_{\alpha;ij}P_{ij}$. Since we neglect phonon-phonon interactions, the lattice distortion energy is exactly calculable as
\begin{equation}
    \delta E[P]=-\sum_{\alpha}\frac{1}{\omega_\alpha}\left|\sum_{ij}\Gamma_{\alpha;ij}P_{ij}\right|^2,
\end{equation}
corresponding to an electronic `Hartree-Fock' Hamiltonian
\begin{equation}
    \hat{H}^\text{HF}[P]=-\sum_{ij}\sum_\alpha \frac{2}{\omega_\alpha}\left(\sum_{kl}\Gamma^*_{\alpha;kl}P^*_{kl}\right)\Gamma_{\alpha;ij}\hat{d}^\dagger_i \hat{d}_j.
\end{equation}
Note the factor of 2 which is typical of terms (like the usual electron interactions) that are quadratic in projectors, and that there is no `Fock' term. The lattice distortion energy is recovered with
\begin{equation}
    \delta E[P]=\frac{1}{2}\sum_{ij}H^\text{HF}_{ij}[P]P_{ij}.
\end{equation}

\section{Additional results}

\begin{table*}
\centering
\newcommand{\colskip}{\hskip 0.15in}
\renewcommand{\arraystretch}{1.15}
\begin{tabular}{
l @{\hskip 0.3in} 
c @{\colskip} 
c @{\colskip} 
c  }\toprule[1.3pt]\addlinespace[0.3em]
Phase & 
$|\nu|$ & 
$\ket{\psi}$ & 
$Q$ 
\\ \midrule
KIVC & 0 & $\left(\ket{KA}+\ket{\bar{K}B})(\ket{KB}-\ket{\bar{K}A}\right)$ & $\tau_x\sigma_y$ \\
TIVC & 0 & $\left(\ket{KA}+\ket{\bar{K}B})(\ket{KB}+\ket{\bar{K}A}\right)$ & $\tau_x\sigma_x$  \\
VH & 0 & $\ket{KA}\ket{\bar{K}A}$ & $\sigma_z$ \\
KIVC & 2 & $\left(\ket{KA\uparrow}+\ket{\bar{K}B\uparrow})(\ket{KB\uparrow}-\ket{\bar{K}A\uparrow}\right)$ & $P_\uparrow\tau_x\sigma_y-P_\downarrow$ \\
TIVC-SP & 2 & $\left(\ket{KA\uparrow}+\ket{\bar{K}B\uparrow})(\ket{KB\uparrow}+\ket{\bar{K}A\uparrow}\right)$ & $P_\uparrow\tau_x\sigma_x-P_\downarrow$ \\
TIVC-QSH & 2 & $\left(\ket{KA\uparrow}+\ket{\bar{K}B\uparrow})(\ket{KB\downarrow}+\ket{\bar{K}A\downarrow}\right)$ & $P_\uparrow\begin{pmatrix}0&0&0&1\\0&-1&0&0\\0&0&-1&0\\1&0&0&0\end{pmatrix}+P_\downarrow\begin{pmatrix}-1&0&0&0\\0&0&1&0\\0&1&0&0\\0&0&0&-1\end{pmatrix}$ \\
IVC-QAH & 2 & $\left(\ket{KA\uparrow}+\ket{\bar{K}B\uparrow})(\ket{KA\downarrow}+\ket{\bar{K}B\downarrow}\right)$ & $s_0\begin{pmatrix}0&0&0&1\\0&-1&0&0\\0&0&-1&0\\1&0&0&0\end{pmatrix}$
 \\
VH & 2 & $\ket{KA\uparrow}\ket{\bar{K}A\uparrow}$ & $P_\uparrow\sigma_z - P_\downarrow$
\\\bottomrule[1.3pt]
\end{tabular}
\caption{\textbf{$Q$-matrices for valley-unpolarized strong-coupling states at even integer filling.} For neutrality, the state in the $\ket{\psi}$ column is repeated for both spin projections. We have used $U_V(1)$ to fix the IVC angle. $P_\uparrow$ $(P_\downarrow)$ is a projector onto spin up (down).\label{tab:app:even}}
\end{table*}

Tab.~\ref{tab:app:even} shows the $Q$-matrix representation for the strong-coupling states shown in Tab.~I in the main text.

\begin{figure*}[t!]
    \centering
    \includegraphics[width=0.85\linewidth]{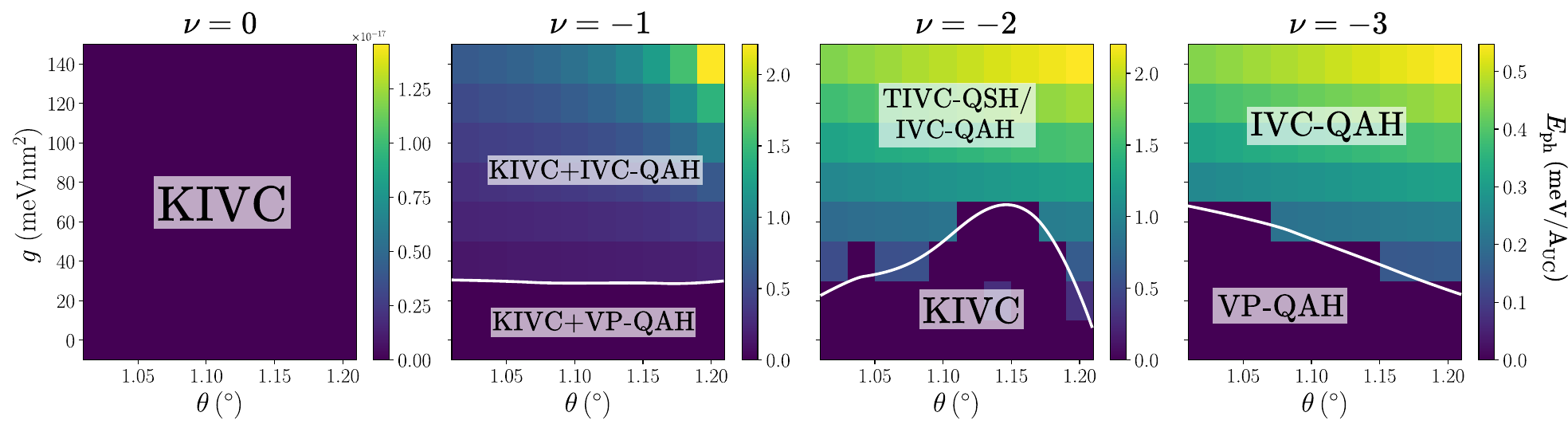}
    \caption{\textbf{Phase diagrams at integer $\nu$ in the graphene subtraction scheme.} Color shows the phonon 
    distortion energy $E_{\text{ph}}$; a non-zero value implies Kekul\'e charge density order. White lines show approximate phase boundaries. Phase diagrams are for $w_\text{AA}=60\,\text{meV},w_\text{AB}=110\,\text{meV}$, and using the graphene subtraction scheme. [QAH: quantized anomalous Hall, KIVC: Kramers intervalley coherent, TIVC: time-reversal intervalley coherent, VP: valley polarized, QSH: quantum spin Hall]}
    \label{fig:app:all_phase_diagram_graphene}
\end{figure*}

The $g$ vs $\theta$ phase diagrams in the main text were computed using the `graphene' subtraction scheme where the electron density for the (screened) Coulomb interactions is measured relative to the density of isolated graphene layers at neutrality. A subtraction scheme is necessary to prevent double-counting of the interactions (otherwise, e.g.~the Hartree-renormalized bands at charge neutrality would be heavily particle-hole asymmetric). In this scheme, the renormalized bandwidth does not vanish, even at the magic angle --- $J$ is a monotonic function of angle. The phase diagrams for all integer $\nu\leq0$ are shown in Fig.~\ref{fig:app:all_phase_diagram_graphene}. Notably, the TIVC does not appear at neutrality for the parameters shown, which can be understood in the NLSM limit from the fact that the TIVC cannot avoid facing the $J$ penalty. At $|\nu|=1$, the ground state in the absence of EPC can be thought of as a KIVC for the (say) up spins, and a VP-QAH for the down spins. With finite $g$, the down spins transition to the IVC-QAH. However the up spins cannot easily transition to the TIVC because of the $J$ penalty. 

\begin{figure*}[t!]
    \centering
    \includegraphics[width=0.85\linewidth]{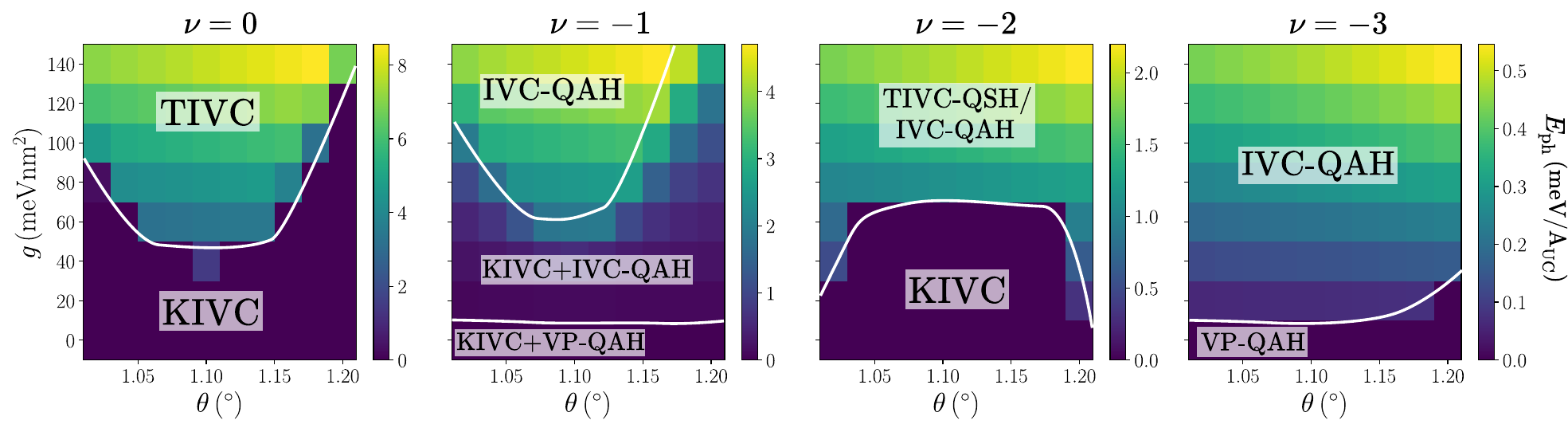}
    \caption{\textbf{Phase diagrams at integer $\nu$ in the average subtraction scheme.} Color shows the lattice 
    distortion energy $E_{\text{ph}}$; a non-zero value implies Kekul\'e charge density order. White lines show approximate phase boundaries. Phase diagrams are for $w_\text{AA}=60\,\text{meV},w_\text{AB}=110\,\text{meV}$, and using the graphene subtraction scheme. [QAH: quantized anomalous Hall, KIVC: Kramers intervalley coherent, TIVC: time-reversal intervalley coherent, VP: valley polarized, QSH: quantum spin Hall]}
    \label{fig:app:all_phase_diagram_average}
\end{figure*}

\begin{figure}[t]
    \centering
    \includegraphics[width=0.5\linewidth]{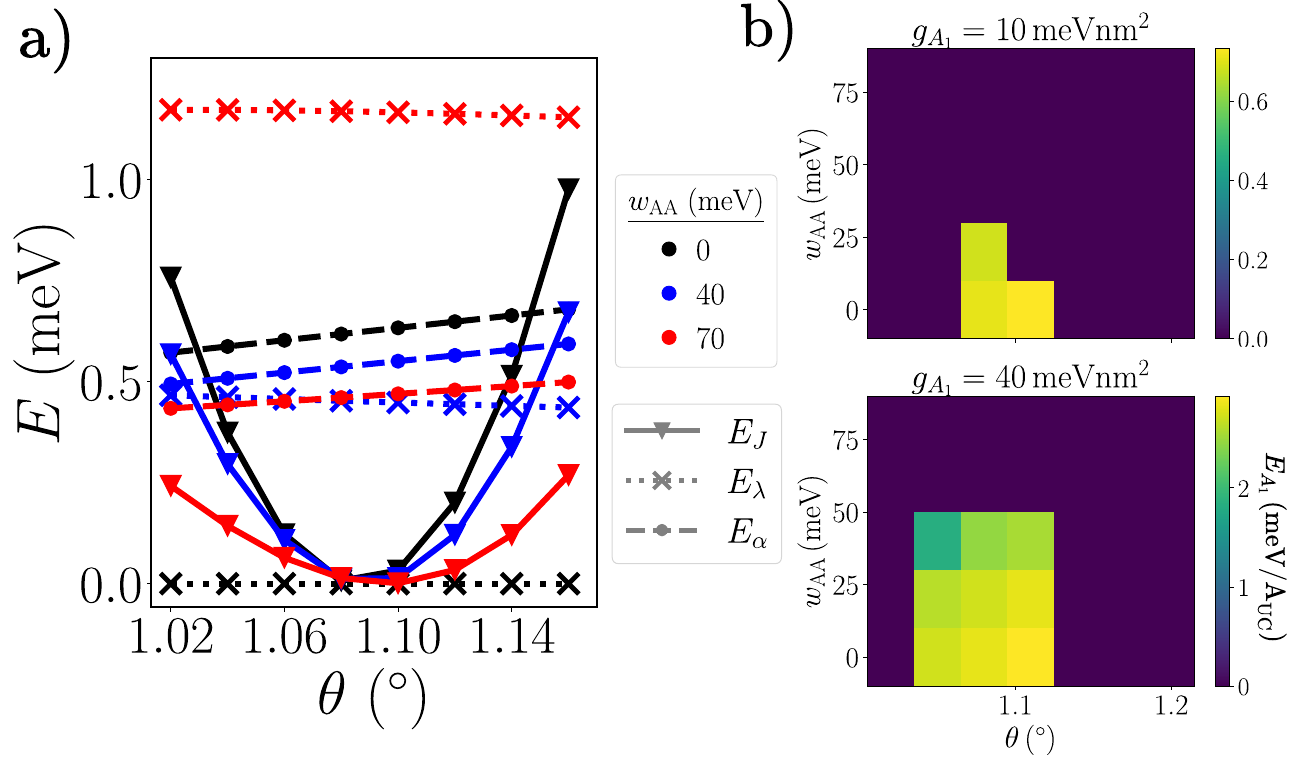}
    \caption{\textbf{Strong-coupling limit  (average scheme).} a) Parameters of the strong-coupling sigma model (energies are per moir\'e unit cell) for $g=70\,$meVnm$^{2}$. $E_\alpha=\alpha A_{\text{UC}}$ (dashed circles) measures twice the lattice distortion energy per unit cell for a single intervalley coherent Chern band. b) Phase diagrams showing lattice distortion energy at $\nu=0$ in the average scheme, where $J=\lambda=0$ in the chiral-flat limit $w_\text{AA}=0,\theta\simeq 1.08^\circ$.}
    \label{fig:app:Jlamb}
\end{figure}

An alternative choice is the `average' subtraction scheme (Fig.~\ref{fig:app:all_phase_diagram_average}), where the electron density for the (screened) Coulomb interactions is measured relative to the infinite temperature density matrix of the central bands. One feature of the average scheme is that at the magic angle in the chiral limit, the interaction-renormalized bandwidth exactly vanishes, such that the superexchange scale $J\sim \frac{t^2}{U}$ is zero. Fig.~\ref{fig:app:Jlamb}a shows the corresponding scales $J,\lambda,\alpha$, and Fig.~\ref{fig:app:Jlamb}b shows that the TIVC at $\nu=0$ in the chiral-flat limit emerges as soon as $g$ is finite. However, a realistic subtraction scheme is unlikely to be so fine-tuned. The corresponding phase diagrams are shown in Fig.~\ref{fig:app:all_phase_diagram_average}. Note the difference with the graphene scheme for $|\nu|=0,1$. On the other hand, the phase diagrams at $|\nu|=2$ and $|\nu|=3$ are qualitatively unchanged since all phases satisfy the $J$-term. 

\end{appendix}

\end{document}